\documentclass[10pt,letterpaper]{article}
\usepackage{opex3}

\usepackage{color}
\usepackage{amsmath}

\newcommand{\NYVO}{Nd:YVO$_4$~}

\usepackage{upgreek} 

\usepackage[backref=page]{hyperref}
\hypersetup{
		colorlinks=true,	      
	  linkcolor=blue,          
    citecolor=blue,        
    urlcolor=blue,           
}

\begin{document}

\title{Comment on: Investigation of the thermal lens effect of the TGG crystal in high-power frequency-doubled laser with single frequency operation}

\author{Ulrich Eismann}

\address{TOPTICA Photonics AG, Lochhamer Schlag 19, 82166 Graefelfing, Germany}




\begin{abstract}
In their article, the authors of [Opt. Express {\bf 23}, 4981 (2015)] observe step-like behavior and hysteresis of the output power of a diode-end-pumped \NYVO-laser as a function of pump power. While claiming that this behavior is a proof of thermal lensing in the intra-cavity TGG, no direct evidence is given. We will discuss here the validity of their statement, and propose an experimental proof.
\end{abstract}

\bibliographystyle{osajnl}
\bibliography{comment}

In Fig.\,5 of their article, the authors of \cite{Yin2015} show step-like and hysteresis behavior of the output power. This is typical for high-power diode-pumped solid state laser designs, and has been observed in the same material system \NYVO, although at a different emission wavelength\,\cite{Lenhardt2009,Eismann2013}. The reason is the crossing of stability boundaries of the eigenmodes of the laser resonator when intracavity thermal lenses are varying. 

In a high-power solid-state laser design, the cold cavity may be unstable. When increasing pump ower, the thermal lens created in the active medium will eventually lead to a stable resonator, causing a step-like increase of the output power at threshold. Furthermore, the thermal lensing of a laser crystal can change significantly depending on whether or not lasing occurs\footnote{This is caused by different pathways in the deexcitation process of the active ions, resulting in different thermal lensing contributions per absorbed pump photon.}. Such, the resonator may remain stable, and the laser on, when reducing the pump power to values smaller than the threshold value. This is what is referred to as hysteresis here, and similar behavior was observed in \cite{Yin2015}.
In the given configuration, the ratio of the thermal lens focal lengths under lasing ($f_{\rm L}$) and non-lasing ($f_{\rm NL}$) conditions gives rise to hysteresis in the output power when $f_{\rm NL}/f_{\rm L} > 1$. In \NYVO, this is the case for the 1342\,nm transition with $f_{\rm NL}/f_{\rm L} = 1.82$\,\cite{Lenhardt2009,Okida2005}. At the 1064\,nm transition used in the present article, however, this value is $f_{\rm NL}/f_{\rm L} = 0.77 < 1$. This change in $f$ does explain a step in output power, but not the hysteresis observed. As a consequence, it does give an indication of thermal lensing of different intra-cavity elements other than the active medium, caused by absorption of circulating laser power. These elements include the TGG crystal, the waveplate, and the LBO frequency-doubling crystal. 

Since the thermal lens value in a medium of length $l$ can be expressed like $1/f \propto \alpha l$, uncertainty of the absorption coefficient $\alpha$ impose a significant error bar. According to \cite{Lu2015}, the TGG vendor specifies $\alpha = 0.0015$\,cm$^{-1}$. Ref.\,\cite{Khazanov2004} gives a value of $\alpha = 0.005$\,cm$^{-1}$, and the authors of \cite{Yin2015} used $\alpha = 0.0010$\,cm$^{-1}$ for their calculations. This factor of three in literature values raises further doubts about the validity of the statement made in \cite{Yin2015}.
Furthermore, the waveplate can cause a thermal focal length of the same order of magnitude as the TGG, because of possibly higher absorption $\alpha$, lower thermal conductivity, and different thermo-optical coefficients, compensating its shorter length $l$.
Finally, the LBO frequency-doubling crystal has thermo-optic and thermal conduction coefficients which are comparable to TGG. Therefore, thermal lensing can be comparable or stronger, especially when strong focussing, higher-order effects like absorption of the second harmonic light, or second-harmonic induced fundamental absorption (SHIFA) are accounted for. In conclusion, without further study, the observation of the output power behavior alone is not a sufficient proof of the occurence of thermal lensing \emph{in TGG}.

In order to gain deeper insight on the stability of the resonator, an  ABCD-matrix based resonator analysis is performed\,\cite{Chunosov2011}. 
We use the resonator parameters (distances, positions and curvatures) available from \cite{Yin2015}, and furthermore $l_{12} = $135\,mm and $l_{23} = l_{41} =$150\,mm, with angles of incidence of 10$^\circ$ on the mirrors, as can be estimated from Fig.\,1 of the article. We treat the thermal lens powers of \NYVO, TGG and waveplate ($f_{\lambda/2}$) in an additive way, and as a single lens,
\begin{equation}
\frac{1}{f_{final}} = \frac{1}{f_{Nd:YVO_4}} + \frac{1}{f_{TGG}} + \frac{1}{f_{\lambda/2}}
\rm ,
\label{E}
\end{equation}
as justified in \cite{Yin2015}, see Eq.\,(4) therein. We assume another single lens centered inbetween M$_3$ and M$_4$, representing the contribution $f_{\rm LBO}$ of the LBO crystal. We calculate a stability diagram of the resonator for which both $f_{final}$ and $f_{\rm LBO}$ are varied, see Fig.\,1. With the definition used here, stability is achieved for $0 \leq  S \leq 1$, where $S = 1-((A+D)/2)^2$ is the stability parameter and $A+D$ is the trace of the total resonator round-trip matrix. The cold resonator is unstable, and when increasing pump power, $f_{final}$ gets smaller and eventually crosses the edge of stability near $f_{\rm final, crit} = 190$\,mm (vertical dashed red line). It is remarkable that the LBO has no significant contribution to the resonator stability in our parameter range.

Without the onset of lasing, the thermal lens in the \NYVO  can be calculated to be $f_{final} = f_{Nd:YVO_4} = 180\,{\rm mm} < f_{\rm final, crit}$, with the values from \cite{Lenhardt2009}, and a pump spot radius of 530\,$\upmu$m\,\cite{Yin2015} (NL region in Fig.\,1). Here, the pump power was taken to be 80\,W, corresponding to where the step in output power was observed. However, when lasing starts, lensing in the \NYVO is reduced and the resonator becomes instable with $f_{Nd:YVO_4} = 240\,{\rm mm} > f_{\rm final, crit}$ alone (L region in Fig.\,1). Therefore, only a further contribution to $f_{final}$, Eq.\,\eqref{E}, explains stability in this regime of operation. Thermal lensing of one of the intra-cavity elements must be the reason, and would lead to hysteresis. Without further explanation, the authors of \cite{Yin2015} claim to have obtained an experimental value of $f_{TGG} = $379\,mm, yielding $f_{final} = $147\,mm with the \NYVO contribution as before, and no thermal lensing in the waveplate (L + T region in Fig.\,1). Under these circumstances, the resonator is stable, and would show hysteresis in the output power, because with Eq.\,(1) we have an effective $f_{\rm NL}/f_{\rm L} > 1$. 
However, the contributions from TGG and waveplate remain indistinguishable because of their additive nature. 
Therefore, we found no proof, but merely an indication of the TGG being at the origin of the behavior observed. An important experimental validity check would be to measure the value of $f_{TGG}$ directly. Removal of the magnet of the TGG, and the waveplate, and observation of the behavior of the resulting non-unidirectional laser with a similar value of intracavity power would be an alternative approach. A direct proof of TGG thermal lensing causing laser resonator stability and hysteresis would be to build an appropriate and simpler cavity, consisting of \NYVO and TGG as sole intra-cavity elements.

In conclusion, the observed behavior can be explained  by thermal lensing inside the TGG crystal, and the waveplate. Both effects are indistinguishable based on the currently available information. Without presenting a direct proof, the results stronlgy indicate thermal lensing inside the TGG, and are of great importance for the design of similar lasers.

\begin{figure}[]
\centering\includegraphics[width=\linewidth]{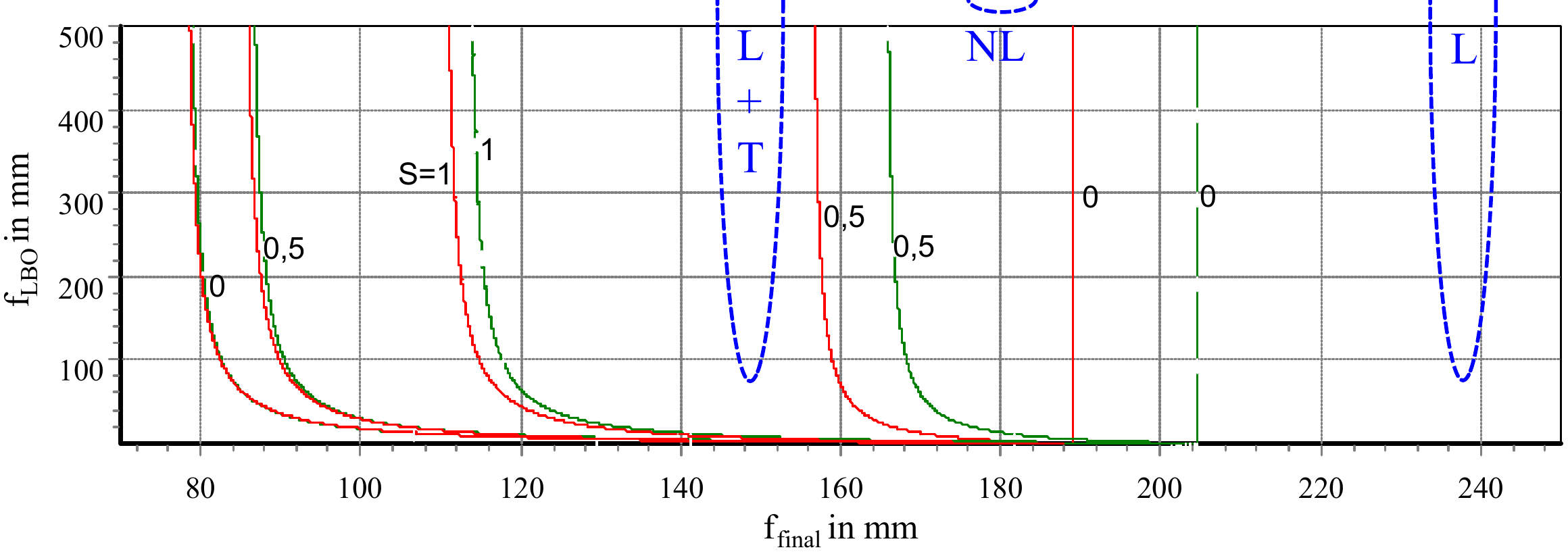}
\caption{The stability diagram of the laser cavity discussed here. The contour plot shows the edges of stability ($S = 0$ in this case) for transversal (green solid line) and sagittal plane (red dashed line). Starting from large values of $f_{final}$ for small pump powers, the cavity becomes stable around $f_{final} = 190$\,mm. The dashed regions indicate typical values for both lenses in the non-lasing case (NL), and the lasing case with (L + T) and without (L) additional thermal lenses inside the laser resonator.}
\label{f:laser}
\end{figure}

\section*{Acknowledgements}
The author wishes to thank A. Bergschneider for helpful comments on the manuscript.

\end{document}